\documentclass[prb,superscriptaddress,twocolumn,showpacs,aps]{revtex4}

\usepackage{graphicx}
\usepackage{dcolumn}
\usepackage{amsmath}

\begin{document}

\title{Neutron scattering study of crystal fields in CeRhIn$_{5}$}
\author{A. D. Christianson}
\affiliation{Los Alamos National Laboratory, Los Alamos NM 87545}
\affiliation{Colorado State University, Fort Collins CO 80523}
\email{achristianson@lanl.gov}
\author{J. M. Lawrence}
\affiliation{University of California, Irvine, CA 92697}
\author{P. G. Pagliuso}
\author{N.O. Moreno}
\author{J. L. Sarrao}
\author{J. D. Thompson}
\affiliation{Los Alamos National Laboratory, Los Alamos NM 87545}
\author{P. S. Riseborough}
\affiliation{Temple University, Philadelphia, PA 19122}
\author{S. Kern}
\affiliation{Colorado State University, Fort Collins CO 80523}
\author{E. A. Goremychkin}
\affiliation{Argonne National Laboratory, Argonne IL 60439}
\author{A. H. Lacerda}
\affiliation{Los Alamos National Laboratory, Los Alamos NM 87545}

\date{\today}

\begin{abstract}
Neutron scattering results for the tetragonal compound
CeRhIn$_{5}$ give evidence for two crystal field (CF) excitations
at 6.9 and 23.6 meV. The scattering can be fit assuming a set of
CF parameters B$^{0}_{2}$ = -1.03 meV, B$^{0}_{4}$ = 0.044 meV and
B$^{4}_{4}$ = 0.122 meV.  To compare our results to previous work,
we calculate the susceptibility and specific heat for this CF
scheme, including a molecular field term $\lambda = $35 mol/emu to
account for the Kondo effect. We also include a calculation based
on these CF parameters that uses the non-crossing approximation to
the Anderson model to estimate the effect of Kondo physics on the
susceptibility, specific heat and neutron linewidths.
\end{abstract}


\maketitle

CeRhIn$_{5}$ crystallizes in the same tetragonal HoCoGa$_{5}$
structure as the heavy fermion superconductors CeIrIn$_{5}$ and
CeCoIn$_{5}$\cite{hegger,p1,p2}. At ambient pressure CeRhIn$_{5}$
undergoes a transition to an antiferromagnetic (AF) state at
$T_{N} =$ 3.8 K\cite{hegger,Bao}. With application of hydrostatic
pressure the N\'{e}el temperature remains essentially constant
until antiferromagnetism disappears and superconductivity appears
at pressures above 15 kbar\cite{hegger}. Recently, Pagliuso
\textit{et al.}\cite{Pagliuso} have suggested the importance of CF
splitting to the ground state properties of the CeMIn$_{5}$ family
of heavy fermion superconductors, underscoring the fact that the
ultimate ground state achieved by a particular member of the
family must grow out of the ground state crystal field doublet.
Thus a careful determination of both the CF splitting and
wavefunctions is important. To that end we have begun to directly
probe the CF energy level splitting in the CeMIn$_{5}$ family
using inelastic (IE) neutron scattering. The first step in our
investigations has been determining the crystal field level scheme
in CeRhIn$_{5}$.

\begin{figure}[t]
\includegraphics[width=3.3in]{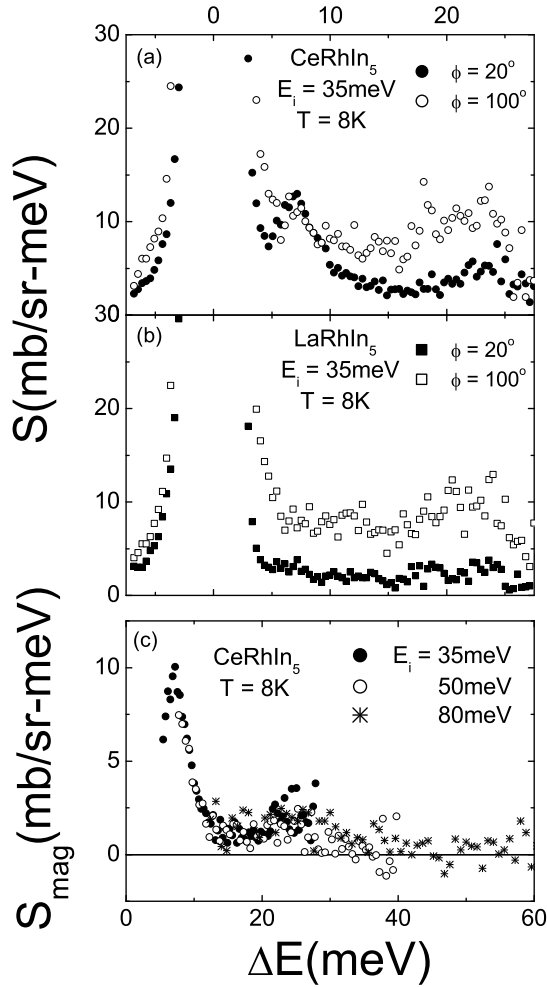}
\caption{Neutron energy spectra of (a) CeRhIn$_{5}$ and (b)
LaRhIn$_{5}$ at an initial energy E$_{i}$ = 35 meV, at 8 K and for
two mean scattering angles, 20$^{\circ}$ and 100$^{\circ}$.  The
data have been corrected for neutron absorption and the scattering
from the sample holder has been subtracted from the data.  (c) The
$Q$=0 magnetic scattering, determined as described in the text, in
CeRhIn$_{5}$ at 8K and for three incident energies $E_{i}$.
\label{fig1}}
\end{figure}

In CeRhIn$_{5}$, as in the other members of CeMIn$_{5}$ family,
the crystal field Hamiltonian in tetragonal symmetry can be
written
\begin{equation*}
H_{CF}=B_{2}^{0}O_{2}^{0}+B_{4}^{0}O_{4}^{0}+B_{4}^{4}O_{4}^{4}
\end{equation*}
where $O_{l}^{m}$ and $B_{l}^{m}$ are the Stevens operators and CF
parameters respectively.  The Ce$^{3+}$ $J=5/2$ wavefunction
splits into three doublets, $\Gamma _{7}^{(1)}=\{\alpha \lbrack
\pm 5/2\rangle +\beta \lbrack \mp 3/2\rangle \}$, $\Gamma
_{7}^{(2)}=\{\beta \lbrack \pm 5/2\rangle -\alpha \lbrack \mp
3/2\rangle \}$ and $\Gamma _{6}=[\pm 1/2\rangle $\cite{Pagliuso}.
\ An analysis of susceptibility and thermal expansion
results\cite{Takeuchi} suggested crystal field levels $\Gamma
_{7}^{(2)}$, $\Gamma _{7}^{(1)}$ and $\Gamma _{6}$ at $E$=0, 5.86
meV (68 K) and 28.43 meV (300 K) respectively, with $\beta=0.969$
(yielding a nearly pure $\lbrack \pm 5/2\rangle$ ground state).  A
subsequent study\cite{Pagliuso} based on an analysis of the
susceptibility and specific heat suggested a similar scheme, but
with splittings 6 and 12 meV (70 and 140 K). In this paper we
report the results of an analysis of neutron scattering data for
CeRhIn$_{5}$ which indicate that these initial estimates are
nearly correct; our results have somewhat different values for the
splittings and a smaller value for the mixing parameter $\beta$,
i.e., a greater admixture of $\lbrack \mp 3/2\rangle$ into the
$\lbrack \pm 5/2\rangle$ ground state. To assist in comparison of
our results to those of Pagliuso \textit{et al.}\cite{Pagliuso}
and Takeuchi \textit{et al.}\cite{Takeuchi}, we report
calculations of the specific heat and magnetic susceptibility
based on our CF parameters which include the Kondo effect in an ad
hoc manner similar to those of refs. 5 and 6.  We also present
more sophisticated calculations that employ the non-crossing
approximation (NCA)\cite{BCW,crossover} to the Anderson model in
order to estimate the effect of Kondo spin fluctuations on the
susceptibility, specific heat and IE neutron spectra.

\begin{figure}[two]
\includegraphics[width=3.3in]{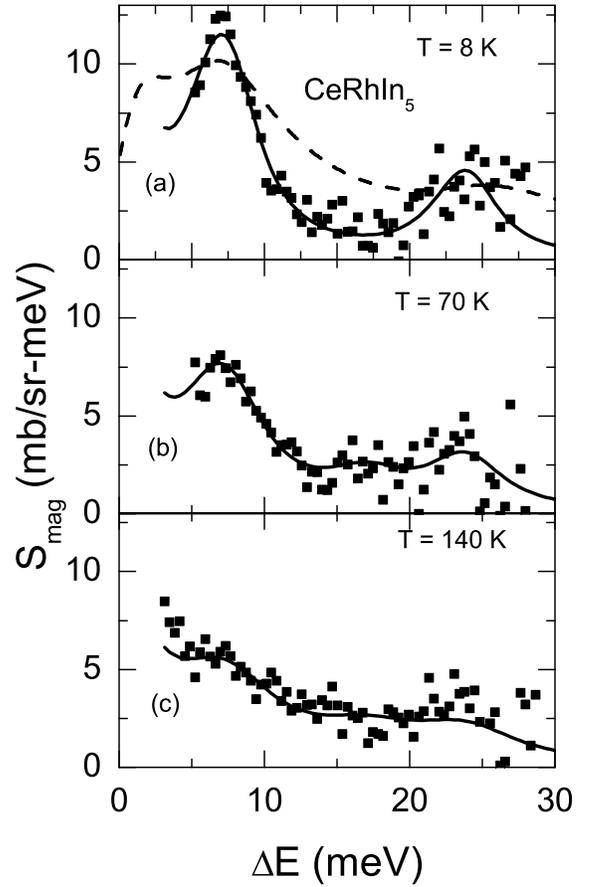}
\caption{Temperature dependence of the magnetic part of the IE
neutron scattering response of CeRhIn$_{5}$ for $E_{i}$ = 35 meV.
The scattering dependence due to the Ce$^{3+}$ form factor has
been removed as in the previous figure.  The data at all three
temperatures ((a) 8 K, (b) 70 K, and (c) 140 K) have been fit
simultaneously (solid lines) with a least squares fitting routine
to determine the crystal field parameters.  The results of the
fitting parameters including the crystal field parameters are
displayed in table \ref{table1}.  We have included in (a) the
results of the NCA calculation (dashed line).}
\end{figure}

Large high quality single crystals of CeRhIn$_{5}$ and
LaRhIn$_{5}$ were obtained using the flux-growth
method\cite{hegger}. For the magnetic susceptibility and specific
heat measurements, single crystals were carefully prepared which
were free of residual In flux; in the case of the neutron
scattering measurements, $\sim$50 g of single crystals for both
CeRhIn$_{5}$ and LaRhIn$_{5}$ were powdered.
 The neutron scattering experiments were performed in time-of-flight
 mode using LRMECS at IPNS (Argonne National Laboratory) with experimental
 conditions that were similar to that of an earlier report\cite{Lawrence}.
A key problem in our investigations was the high neutron
absorption of both In and Rh. In initial experiments the standard
LRMECS sample holder was used; however, in subsequent experiments
a new sample holder was employed which was designed to maintain a
more uniform sample thickness than the standard holder, thus
allowing for a more accurate absorption correction.
 Neutron scattering spectra were collected for several different
incident energies ($E_{i}$) and temperatures between 8 and 140 K
with counting times ranging from 24 to 48 hours. To improve
statistics, we were able to take advantage of the nondispersive
nature of the CF scattering and group detectors into three bins
with mean scattering angle 20$^{o}$ (low $Q$), 60$^{o}$ and
100$^{o}$ (high $Q$). A Vanadium standard was utilized to put the
scattering on an absolute scale.

Data for CeRhIn$_{5}$ and LaRhIn$_{5}$ (measured to help identify
the nonmagnetic scattering in CeRhIn$_{5}$) at 8 K and $E_{i} = $
35 meV for low and high Q are shown in Fig. \ref{fig1}.  The data
were corrected for absorption assuming a uniformly thick
flat-plate sample and the scattering due to the empty sample
holder was then subtracted from the data.  Direct comparison of
low angle scattering (where magnetic scattering is strongest) for
CeRhIn$_{5}$ (Fig. \ref{fig1}a) and LaRhIn$_{5}$ (Fig.
\ref{fig1}b) shows two additional peaks near 7 and 23 meV. In
particular, we determine the nonmagnetic scattering in
CeRhIn$_{5}$ in two ways: 1) By using the expression
$S_{mag}(20^{\circ})=S(Ce,20^{\circ})-f S(La,20^{\circ})$ where we
choose the factor $f$ as the ratio (0.75) of the total scattering
cross-sections $\sigma$(CeRhIn$_{5}$)/$\sigma$(LaRhIn$_{5}$). 2)
By determining the ratio $R=S(La,100^{\circ})/S(La,20^{\circ})$
for scaling the high angle nonmagnetic scattering to low
angle\cite{Goremychkin}. Excellent agreement with 1) is obtained
using $S_{mag}(20^{\circ})=S(Ce,20^{\circ})-F S(Ce,100^{\circ})/R$
with inclusion of an additional factor F = 1.33 to account for the
difference in Q-scaling of the La and Ce compounds. The value of F
is similar to the one used in a recent study of
YbXCu$_{4}$\cite{crossover}; it can be justified on the basis that
for high angle scattering the data are predominantly
single-phonon, proportional to $\sigma$, while the low angle
scattering contains a significant contribution from multiple
scattering (one elastic and one phonon) proportional to
$\sigma^{2}$, so that the cross section does not cancel in the
ratio. Results of this analysis for three different $E_{i}$ are
shown in Fig. 1c.  The dependence of the scattering on the
Ce$^{3+}$ form factor has been removed in this plot, so the data
represent the $Q = $ 0 scattering with the assumption that the
crystal fields are in fact purely local and uncoupled entities.
The data have been truncated below 0.15$E_{i}$ (where the elastic
line dominates the scattering) and above 0.8$E_{i}$, where
statistics are small due to the $k_{f}/k_{i}$ factor. Good
agreement is evident for data taken at three different $E_{i}$,
with all data sets displaying magnetic excitations at
approximately 7 and 24 meV.

\begin{table}[tbp]
\caption{ Crystal field parameters $B_{l}^{m}$, splittings and
Lorentzian halfwidths $\Gamma$ of the IE excitations at four
temperatures for CeRhIn$_{5}$ and the wave function mixing
parameter $\beta$. The units of all quantities (except for
$\beta$, which is unitless) are meV. The reduced Chi-square for
the fit was $\chi^{2} = 0.69$} \label{table1}
\begin{tabular}{lll}
$B_{2}^{0}$ &$B_{4}^{0}$ &$B_{4}^{4}$\\
 \tableline -1.03$\pm$0.02 & 0.044$\pm$0.001 &
0.122$\pm$0.003 \\
\tableline  & &\\
$E(\Gamma_{7}^{1})$ &$E(\Gamma_{6})$ &$\beta$\\
\tableline 6.9$\pm0.3$ & 23.6$\pm0.5$ & 0.80$\pm$0.02\\
\tableline  & &\\
$\Gamma (8K)$ & $\Gamma (70K)$  & $\Gamma (140K)$
\\
 \tableline  2.3$\pm$0.1 & 2.9$\pm$0.2  & 4.2$\pm$0.4\\
\tableline  \\
\end{tabular}
\end{table}

In Fig. 2 we plot the $Q = $ 0 (form factor removed) magnetic
scattering (method 1), determined at $E_{i}= $ 35 meV, for three
different temperatures.  We have performed a simultaneous least
squares fit to four datasets (8 K, 70 K and 140 K at $E_{i} = $ 35
meV and 8 K at $E_{i} = $ 80 meV) to determine the CF parameters.
The fit includes the effects of instrumental resolution. Variables
of the fit include $B_{2}^{0}, B_{4}^{0}, B_{4}^{4}$ and an
overall scale factor (which four parameters were constrained to
the same values for all datasets) and the Lorentzian halfwidth
$\Gamma$ of the IE excitations which was allowed to vary with
temperature. (We constrained the quasi-elastic (QE) halfwidth to
1/2 $\Gamma$.) Results of the fit are shown in Table \ref{table1}
and plotted in Fig. 2.

To compare our results to those of Pagliuso \textit{et
al.}\cite{Pagliuso} and Takeuchi \textit{et al.}\cite{Takeuchi},
we have calculated the susceptibility and specific heat (Fig. 3).
The susceptibility includes a positive molecular field
contribution $\lambda = $ 35 mol/emu where $\lambda$ represents
contributions to $1/\chi$ from AF and Kondo fluctuations.  At high
T these contribute to $1/\chi$ as $(T_{K}+T_{N})/C_{J}$; with
$C_{J}$ = 0.807 emu-K/mol for $J=5/2$ and $T_{N}$ = 3.8 K this
gives $T_{K} \sim$ 25 K. We note that this value of $k_{B}T_{K}$
is similar to the width of the 7 meV IE excitation at 8 K.  The
calculation for the specific heat contains both a Schottky term
due to the excited levels and a Kondo doublet term\cite{Rajan}
with $T_{K} = $ 25 K for the ground state level, which puts the
calculated specific heat in the range 20-50 K in better agreement
with measured value -- without this, the calculated value due only
to the Schottky contribution is smaller by a factor of 0.8. We
have not attempted to fit for the effects on $C_{mag}$ and $\chi$
of the AF transition at 3.8 K.

\begin{figure}[where]
\includegraphics[width=3.2in]{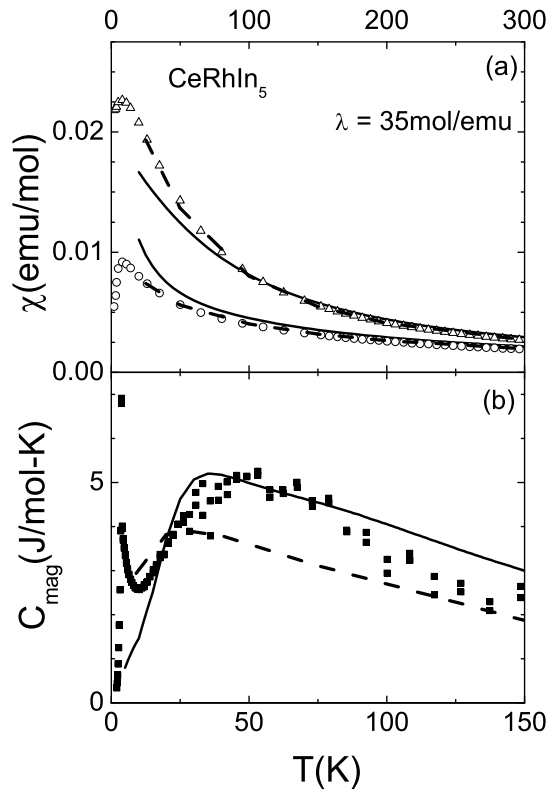}
\caption{ (a) Measured anisotropic susceptibilities $\chi^{zz}$
(triangles) and $\chi^{xx}$ (circles) for CeRhIn$_{5}$ compared to
the value calculated for the CF parameters of Table \ref{table1}
with a molecular field contribution $\lambda =$ 35 mol/emu (solid
lines) and compared to the results of the NCA calculation (dashed
lines). (b) Magnetic specific heat compared to the value
calculated for a Schottky contribution from the excited levels and
a Kondo contribution from the ground state doublet (solid line)
and to the results of the NCA calculation (dashed line).}
\end{figure}

A more sophisticated way to include Kondo spin fluctuations is
through calculation for the Anderson model.  We present results
obtained using the non-crossing approximation
(NCA)\cite{BCW,crossover}. We have used a Gaussian background band
with density of states $N(\varepsilon
)=e^{-(\varepsilon/W)^{2}}/\surd \pi W$ with $W$ = 3 eV and we set
the 4$f$ level position at $E_{f} = 2$ eV and the spin-orbit
splitting at $E_{so} = 0.273$ eV, which are standard values for
Ce. Since the Kondo physics renormalizes the CF levels upward by
an amount approximately equal to the Kondo temperature the bare
level energies were chosen to be $E_{b}=5.3$ meV and $E_{c}=23$
meV which are smaller than the measured level energies. The mixing
parameter $\beta=0.80$ was chosen to be similar to that obtained
in Table 1. The hybridization was then varied until a good fit to
the anisotropic susceptibility was obtained for $V=0.4665$ eV. The
results for $S_{mag}$,  $\chi$ and $C_{mag}$ are given in Figs. 2a
and 3.

We now turn to discussion of the effect of systematic errors on
our conclusions.  As mentioned previously, the neutron absorption
of In and Rh is an important consideration. Comparison of the data
for two different sample holders (which exhibited small
differences in sample thickness and distribution) indicated
similar results, augmenting our belief that the absorbtion
correction employed is correct. If the nonmagnetic background
subtraction is varied by varying $f$ (method 1) or $F/R$ (method
2), the scattering at 7 meV is relatively unaffected but the
strength of the 24 meV scattering, and hence $\beta$, is affected
somewhat. Given the good consistency between results at different
$E_{i}$ and $T$ we think that our CF scheme is basically correct.
We were unable to observe quasi-elastic (QE) scattering, due to
the requirement that to obtain the resolution necessary a small
E$_{i}$ is required which causes the effects of neutron
absorption, which varies as 1/$\surd E$, to become large. In our
fits we constrained the QE halfwidth to half the value of the IE
width to prevent proliferation of fit parameters. Constraining to
other values (e.g. $\Gamma_{QE}=\Gamma_{IE}$) leads only to minor
variation in the final fits.

Our fits to $\chi$ using the CF parameters plus a molecular field
term are not as good as those of Takeuchi \textit{et
al.}\cite{Takeuchi} or Pagliuso \textit{et al.}\cite{Pagliuso}.
However, their fits use a value of $\beta$ very close to unity,
indicating essentially no $[\mp 3/2\rangle$ admixture into $[\pm
5/2\rangle$ ground state.  In this case there would be no
observable amplitude for the $\Delta m_{z}=1$ transition to $[\pm
1/2\rangle$ state at 24 meV. This cannot be correct as we clearly
observe this transition in the neutron scattering data.  A
possible reason that our fits are not as good as those of refs. 5
and 6 is that we do not include the effect of exchange anisotropy,
which should only be important below 20 K. Such anisotropy can be
mimicked as in Pagliuso \textit{et al.}\cite{Pagliuso} through
inclusion of an anisotropic mean field parameter, which we have
chosen not to do for simplicity.

On the other hand, the NCA calculations based on our CF scheme and
a Kondo temperature of order 25 K does an excellent job
reproducing $\chi$.  However, it overestimates the width of the 7
meV excitation as seen in Fig. 2a and underestimates the
temperature of the peak in the specific heat (Fig. 3b). These
deviations from the data may reflect the fact that we have neither
included antiferromagnetic exchange, exchange anisotropy nor
anisotropic hybridization (i.e. different hybridization to the
different CF multiplets) in the NCA fits.

In summary, we find a more significant $[\mp 3/2\rangle$ admixture
into $[\pm 5/2\rangle$ ($\beta=0.80$) ground state than found
earlier by Takeuchi \textit{et al.} ($\beta=.969$)\cite{Takeuchi}
or Pagliuso \textit{et al.} ($\beta\sim1$)\cite{Pagliuso}. The
resulting CF level parameters provide reasonable fits to both the
magnetic susceptibility and specific heat with the inclusion of a
mean field parameter and a Kondo doublet respectively. In
addition, NCA fits the susceptibility remarkably well with some
deficiencies in both the specific heat and neutron scattering
linewidths.  Taken together the NCA calculations and the fits to
specific heat and susceptibility all indicate a $T_{K} \sim$ 25 K.
We note that the ordered moment $g \mu _{B} \langle J_{z} \rangle
= $ 0.92$\mu _{B}$ deduced for $\beta =$0.80 is substantially
larger than the value 0.36$\mu_{B}$ needed to fit the diffraction
pattern in the ordered state\cite{Bao}.  For our estimate of
$T_{K}$ the moment is reduced by the Kondo physics at temperatures
$T > T_{N}$.

We acknowledge useful discussions with W. Bao and R.J. McQueeney.
Work at UC Irvine was supported by UCDRD funds provided by the
University of California for the conduct of discretionary research
by the Los Alamos National Laboratory and by the UC/LANL Personnel
Assignment Program. Work at Los Alamos and Argonne was performed
under the auspices of the DOE. The work at Temple University was
supported by the Department of Energy, award DE-Fg02-01ER45827.

\end{document}